\documentclass[aps,pra,onecolumn,floatfix,notitlepage,superscriptaddress]{revtex4-1}

\usepackage{amsmath} 
\usepackage{graphicx} 

\begin{document}

\title{PT-symmetric non-Hermitian quantum many-body system using ultracold atoms in an optical lattice with controlled dissipation}





\author{Yosuke Takasu}
\affiliation{Department of Physics, Graduate School of Science, Kyoto University, Kyoto 606-8502, Japan}
\author{Tomoya Yagami}
\affiliation{Department of Physics, Graduate School of Science, Kyoto University, Kyoto 606-8502, Japan}
\author{Yuto Ashida}
\affiliation{Department of Applied Physics, University of Tokyo, 7-3-1 Hongo, Bunkyo-ku, Tokyo 113-8656, Japan}
\author{Ryusuke Hamazaki}
\affiliation{Department of Physics, The University of Tokyo, 7-3-1 Hongo, Bunkyo-ku, Tokyo 113-0033, Japan}
\affiliation{Nonequilibrium Quantum Statistical Mechanics RIKEN Hakubi Research Team, RIKEN Cluster for Pioneering Research (CPR), RIKEN iTHEMS, Wako, Saitama 351-0198, Japan}
\author{Yoshihito Kuno}
\affiliation{Department of Physics, Graduate School of Science, Kyoto University, Kyoto 606-8502, Japan}
\author{Yoshiro Takahashi}
\affiliation{Department of Physics, Graduate School of Science, Kyoto University, Kyoto 606-8502, Japan}


\begin{abstract}%
We report our realization of a parity-time (PT) symmetric non-Hermitian many-body system using cold atoms with dissipation. After developing a theoretical framework on PT-symmetric many-body systems using ultracold atoms in an optical lattice with controlled dissipation, we describe our experimental setup utilizing one-body atom loss as dissipation with special emphasis on calibration of important system parameters. We discuss loss dynamics observed experimentally.
\end{abstract}


\maketitle

\section{Introduction}
The last two decades have witnessed remarkable developments in studies of out-of-equilibrium dynamics in isolated quantum many-body systems, as mainly promoted by experimental advances in atomic physics \cite{BI08,LDA16}. On another front, it has become possible to study {\it open} many-body physics in a highly controlled manner \cite{NS08,BG13,PYS15,RB16,LHP17,Tomita2017,LJ19,CL19}. Under such nonunitary perturbations acted by an external observer or a Markovian environment, the dynamics of an open quantum system can be described by the quantum trajectory approach \cite{DJ92,HC93,MU92}. With the ability to measure single quanta of many-body systems \cite{BWS09,SJF10,MM15,CLW15,PMF15,OA15,EH15,EGJ15,RY16,AA16}, one should unravel open-system dynamics based on microscopic information, e.g., the number of quantum jumps occurred in a certain time interval \cite{Daley2014,YA18}. Non-Hermitian quantum physics is a useful description to investigate such dynamics conditioned on measurement outcomes. From a broader perspective, we remark that non-Hermitian quantum physics has also been applied to a problem of nuclear and mesoscopic resonances \cite{HF58,NM98}, flux-line depinning  \cite{HH96,IA04}, and quantum transport \cite{GGG15}. In particular, the Feshbach projection approach \cite{HF58} has recently been revisited in the context of correlated electrons \cite{VK17,YT18}. 

Non-Hermitian physics has also attracted considerable interest from other diverse subfields of physics in this decade \cite{EG18,HX2016,Gong2018,LJY19,FZN19%
,Liu2019,Ashida2020%
}; one prominent example is its application to linear dynamics in classical systems. The reason is that one can regard a set of classical wave equations as the one-body Schr{\"o}dinger equation, allowing one to simulate non-Hermitian wave physics in classical setups \cite{AR05,REG07}. More specifically, recent advances in this direction have been sparked by realizations \cite{GA09,RCE10} of the parity-time (PT) symmetric quantum mechanics \cite{Bender1998} with fabricated gain and loss in chip-scale nanophotonic systems. In PT-symmetric systems, spectra can exhibit real-to-complex transitions typically accompanying singularities known as exceptional points \cite{TK80}. While such transitions can be found also for other types of antilinear symmetries \cite{CMB02}, an advantage in the PT symmetry is its physical simplicity, allowing one to ensure the symmetry via spatial modulation of dissipative media~\cite{Ozdemir2019,Peng2014,Peng2014b,Jing2014,Jing2015,Liu2016,Zhang2018,Liu2017,Quijandria2018,Zhang2015}. The subsequent studies
revealed a great potential of exploring non-Hermitian physics in a variety of classical systems such as  mechanical systems \cite{KB17}, electrical circuits \cite{CHL18,IS18,EM19a}, and acoustic or hydrodynamic systems \cite{SC16,ZW18,SK19}. Yet,  these developments have so far been mainly restricted to one-body wave phenomena.

Recently, there have been significant efforts to extend the paradigm of non-Hermitian physics to yet another realm of quantum many-body systems. Already, a number of unconventional many-particle phenomena that have no analogues in Hermitian regimes have been found  \cite{YA17nc,LTE14,MG16,YA16crit,SJP16,LAS18,NM18,GA18,Hamazaki2019,YK19,YT19,LE19,MN19,GC19}. In particular, it has been theoretically proposed to realize a PT-symmetric quantum many-body system in ultracold atoms by using a spatially modulated dissipative optical potential, where the renormalization group (RG) fixed points and RG flows unique to non-Hermitian regimes have been predicted \cite{YA17nc}. Similar anomalous RG flows have also been studied in the Kondo model \cite{LAS18,NM18}. Also by considering a double-well optical lattice, a PT-symmetric Bose condensed system has been theoretically proposed \cite{Dizdarevic2018}. More recently, studies have been extended to nonequilibrium dynamical regimes \cite{YS17,DB192,DB19}, where the role of quantum jumps in time evolution has been elucidated \cite{YA18,YA18therm,NM19e,GC20}.  

The aim of this paper is to report our realization of a PT-symmetric non-Hermitian quantum many-body system using ultracold ytterbium (Yb) atoms in an optical lattice. 
While a highly controllable system of ultracold atoms in an optical lattice is inherently an isolated quantum many-body system, this controllability also enables us to realize open quantum systems by coupling the system to the environment.
We used controlled one-body atom loss as dissipation.
%
Specifically, we investigate loss dynamics of a one-dimensional Bose gas, which is the key for understanding this non-Hermite PT-symmetric system, revealed by our theory.
This paper is organized as follows: In Sec. 2 we describe our newly developed theoretical framework.
Then we describe our experiment with one-body loss in Sec. 3.
Finally, we discuss the obtained results and summarize our work in Sec. 4.
In the Appendix we describe the results for two-body loss.

\section{Theoretical framework}
In this section, we develop a theoretical framework to describe the experimental system considered in the present paper. We note that the formulation described here is equally applicable to both bosonic and fermionic systems.
\subsection{Model}
We consider atoms trapped in an optical lattice created by a far-detuned laser. We then discuss superimposing a dissipative optical lattice created by near-resonant light with a possible phase shift $\phi$. When the shift $\phi$ satisfies a certain condition discussed below, an effective non-Hermitian Hamiltonian in the system can satisfy the PT symmetry in the sense of a passive system. 

Suppose that atoms have the ground $|g\rangle$ and excited $|e\rangle$
states with the atomic frequency $\omega_{0}$, and the excited state
has decay channels to other states with the total decay rate $\Gamma$,
which is much larger than the spontaneous emission rate of $|e\rangle$
to $|g\rangle$. The dynamics can then be described by the many-body
Lindblad equation:
\begin{eqnarray}
\frac{d\hat{\rho}}{dt}  =  -\frac{i}{\hbar}[\hat{H},\hat{\rho}]-\frac{\Gamma}{2}\int\left[\hat{\Psi}_{e}^{\dagger}({\bf x})\hat{\Psi}_{e}({\bf x})\hat{\rho}+\hat{\rho}\hat{\Psi}_{e}^{\dagger}({\bf x})\hat{\Psi}_{e}({\bf x})-2\hat{\Psi}_{e}({\bf x})\hat{\rho}\hat{\Psi}_{e}^{\dagger}({\bf x})\right]d{\bf x},
\end{eqnarray}
where $\hat{\Psi}_{g,e}^\dagger$ ($\hat{\Psi}_{g,e}$) is the creation (annihilation) operator of the ground- or excited-state atom, and $\hat{H}$ is the Hamiltonian of the two-level atoms,
\begin{eqnarray}
\hat{H} & \!=\! & \int d{\bf x}\left[\sum_{i=g,e}\hat{\Psi}_{i}^{\dagger}({\bf x})\hat{H}_{i,{\rm CM}}({\bf x})\hat{\Psi}_{i}({\bf x})\!+\!\hbar\omega_{0}\hat{\Psi}_{e}^{\dagger}({\bf x})\hat{\Psi}_{e}({\bf x})\!+\!\hat{H}_{{\rm I}}({\bf x})\!+\!\hat{H}_{{\rm int}}({\bf x})\right],\\
\hat{H}_{i,{\rm CM}}({\bf x}) &\! =\! & -\frac{\hbar^{2}\nabla^{2}}{2m}\!+\!U_{i}({\bf x})\;\;\;\;\;\;(i=g,e),\\
\hat{H}_{{\rm I}}({\bf x}) & = & -\left({\bf d}\cdot{\bf E}({\bf x},t)\hat{\Psi}_{g}^{\dagger}({\bf x})\hat{\Psi}_{e}({\bf x})+{\rm H.c.}\right),\\
\hat{H}_{{\rm int}}({\bf x}) & = & \frac{u}{2}\hat{\Psi}_{g}^{\dagger}({\bf x})\hat{\Psi}_{g}^{\dagger}({\bf x})\hat{\Psi}_{g}({\bf x})\hat{\Psi}_{g}({\bf x}). \label{eq:u}
\end{eqnarray}
Here, $U_{i}({\bf x})$ are conservative optical potentials created by
far-detuned light,  ${\bf E}({\bf x},t)=2{\bf E}_{0}({\bf x})\cos(\omega_{{\rm L}}t)$ is the electric-field of near-resonant standing-wave light, and ${\bf d}=\langle e|\hat{{\bf d}}|g\rangle$ is the electric dipole moment. 

We then transform to the rotating frame
\begin{equation}
\hat{\tilde{\Psi}}_{e}({\bf x})\equiv e^{i\omega_{{\rm L}}t}\hat{\Psi}_{e}({\bf x}).
\end{equation}
Applying the rotating wave approximation, the Heisenberg
equation of the motion becomes
\begin{eqnarray}
\dot{\hat{\tilde{\Psi}}}_{e}({\bf x}) & = & i\delta\hat{\tilde{\Psi}}_{e}({\bf x})+\frac{i\Omega^{*}({\bf x})}{2}\hat{\Psi}_{g}({\bf x})-\frac{\Gamma}{2}\hat{\tilde{\Psi}}_{e}({\bf x}),\\
\Omega({\bf x}) & \equiv & \frac{2{\bf d}\cdot{\bf E}_{0}({\bf x})}{\hbar},
\end{eqnarray}
where $\delta\equiv\omega_{{\rm L}}-\omega_{0}$ is the detuning frequency.
We assume the condition $|\delta+i\Gamma/2|>\Omega$ and then perform
the adiabatic elimination of the excited state $|e\rangle$:
\begin{eqnarray}
\hat{\tilde{\Psi}}_{e}({\bf x}) & \simeq & -\frac{1}{\delta+\frac{i\Gamma}{2}}\frac{\Omega^{*}({\bf x})}{2}\hat{\Psi}_{g}({\bf x})\\
 & \simeq & \frac{i\Omega^{*}({\bf x})}{\Gamma}\hat{\Psi}_{g}({\bf x}),
\end{eqnarray}
where we use the resonant condition $\delta\ll\Gamma$. The
master equation of the ground-state atoms now reduces to 
\begin{equation}
\frac{d\hat{\rho}}{dt}=-\frac{i}{\hbar}(\hat{H}_{{\rm eff}}\hat{\rho}-\hat{\rho}\hat{H}_{{\rm eff}}^{\dagger})+\int dx\frac{|\Omega({\bf x})|^{2}}{\Gamma}\hat{\Psi}({\bf x})\hat{\rho}\hat{\Psi}^{\dagger}({\bf x}),\label{eq:master}
\end{equation}
where we abbreviate the index $g$ and redefine the mass and the potential
by incorporating the renormalization factor induced by the excited
state. We here also introduce the effective non-Hermitian Hamiltonian
\begin{equation}
\hat{H}_{{\rm eff}}=\int d{\bf x}\hat{\Psi}^{\dagger}({\bf x})\left(-\frac{\hbar^{2}\nabla^{2}}{2m}+U({\bf x})-\frac{i\hbar|\Omega({\bf x})|^{2}}{2\Gamma}\right)\hat{\Psi}({\bf x})+\frac{u}{2}\int d{\bf x}\hat{\Psi}^{\dagger}({\bf x})\hat{\Psi}^{\dagger}({\bf x})\hat{\Psi}({\bf x})\hat{\Psi}({\bf x}).
\end{equation}

For the sake of simplicity, hereafter we set $u=0$ and consider  a one-dimensional (1D) system subject to an off-resonant optical lattice 
\begin{equation}
U(x)=\frac{V_{0}}{2}\cos\left(\frac{2\pi x}{d}\right),
\end{equation}
where $V_{0}$ is the lattice depth and $d=\lambda/2$ is the lattice
constant. We consider superimposing a near-resonant optical lattice having the (shorter) lattice constant $d/2=\lambda/4$ with a phase shift $\phi$. The resulting time-evolution equation is \cite{KSJ98,RS05,YA17nc}
\begin{eqnarray}
\frac{d\hat{\rho}}{dt}&=&-\frac{i}{\hbar}(\hat{H}_{{\rm eff}}\hat{\rho}-\hat{\rho}\hat{H}_{{\rm eff}}^{\dagger})+\int dx\gamma_{0}\left(1+\sin\left(\frac{4\pi x}{d}+\phi\right)\right)\hat{\Psi}(x)\hat{\rho}\hat{\Psi}^{\dagger}(x),\label{eq:lind}\\
\hat{H}_{{\rm eff}} & = & \int dx\hat{\Psi}^{\dagger}(x)\left(-\frac{\hbar^{2}\nabla^{2}}{2m}+V(x)\right)\hat{\Psi}(x)-\frac{i\hbar\gamma_{0}}{2}\hat{N},\label{eq:non-her}\\
V(x) & = & \frac{V_{0}}{2}\left(\cos\left(\frac{2\pi x}{d}\right)-i\gamma\sin\left(\frac{4\pi x}{d}+\phi\right)\right),\\
\gamma & \equiv & \frac{2|d|^{2}\mathcal{E}_{0}^{2}}{\hbar\Gamma V_{0}},\;\;\;\;\;\;\gamma_{0}  \equiv  \frac{V_{0}\gamma}{\hbar}.
\end{eqnarray}

We remark that the effective Hamiltonian $\hat{H}_{\rm eff}$ satisfies the PT symmetry at the phase shift $\phi=0$ aside from a global imaginary constant $-i\hbar \gamma_0 N/2$ because of the relation $V(x)=V^*(-x)$. In this case, the real-to-complex spectral transition occurs at $\gamma_{\rm PT}\simeq 0.25$. Below this threshold $\gamma<\gamma_{\rm PT}$, the PT symmetry is unbroken, i.e., every eigenstate respects the symmetry and the entire spectrum is real. In particular, the band structure associated with the complex potential $V(x)$ exhibits gaps. As we increase $\gamma$, the energy gap between the second and third bands at $k=0$ becomes smaller and closes at $\gamma=\gamma_{\rm PT}$, where two merging eigenstates at $k=0$ coalesce into one, leading to the exceptional point. Above the threshold $\gamma>\gamma_{\rm PT}$, the PT symmetry is broken, i.e., eigenstates around $k=0$ turn to have complex eigenvalues.

\subsection{Relating the non-Hermitian spectrum to loss dynamics }
Signatures of the PT-symmetry breaking in the non-Hermitian Hamiltonian $\hat{H}_{{\rm eff}}$ can be extracted from the dynamics of the density matrix $\hat{\rho}$ obeying the master equation~(\ref{eq:master}). To this end, we focus on the time evolution of the loss rate $L(t)$ of the ground-state atoms:
\begin{equation}
L(t)\equiv-\frac{1}{N(t)}\frac{dN(t)}{dt}=-\frac{1}{N(t)}{\rm Tr}\left[\frac{d\hat{\rho}(t)}{dt}\int dx\hat{\Psi}^{\dagger}(x)\hat{\Psi}(x)\right].
\end{equation}
We then introduce the quantity $\mathcal{L}_{\lambda}(\hat{\rho})$ defined as follows:
\begin{equation}
\mathcal{L}_{\lambda}(\hat{\rho})\equiv\frac{1}{N}{\rm Tr}\left[e^{\frac{i\hat{H}_{{\rm eff}}\lambda}{\hbar}}\hat{\rho}e^{-\frac{i\hat{H}_{{\rm eff}}^{\dagger}\lambda}{\hbar}}\right],\;\;\;N\equiv{\rm Tr}[\hat{N}\hat{\rho}].\label{eq:calL}
\end{equation}
From Eqs. (\ref{eq:non-her}) and (\ref{eq:calL}), one can readily show the following relation:
\begin{equation}
\left.\frac{\partial_{\lambda}\mathcal{L}_{\lambda}(\hat{\rho}(t))}{\partial\lambda}\right|_{\lambda=0}=L(t).\label{eq:pre1}
\end{equation}

To simplify the expression, we employ the spectrum decomposition of the many-body non-Hermitian Hamiltonian:
\begin{eqnarray}
\hat{H}_{{\rm eff}}&=&\sum_{\alpha}\mathcal{E}_{\alpha}|\Psi_{\alpha}\rangle\langle\Phi_{\alpha}|,\\
\mathcal{E}_{\alpha}&=&E_{\alpha}{+}\frac{i\Gamma_{\alpha}}{2}-\frac{i\hbar\gamma_{0} N_\alpha}{2},
\end{eqnarray}
where $\mathcal{E}_{\alpha}$ is a complex eigenvalue with $E_{\alpha}$ ($\Gamma_{\alpha}$) denoting its real (imaginary) part, $N_\alpha$ is a particle number, and $|\Psi_{\alpha}\rangle$ and $|\Phi_{\alpha}\rangle$ are the corresponding right and left
eigenvectors, respectively, which satisfy $\langle\Psi_{\alpha}|\Phi_{\beta}\rangle = \delta_{\alpha\beta}$ and $\sum_{\alpha}|\Psi_{\alpha}\rangle\langle\Phi_{\alpha}|  =  \hat{I}$.
From Eq.~(\ref{eq:pre1}), one can relate
the loss dynamics to the spectrum of the non-Hermitian Hamiltonian via
\begin{eqnarray}
L(t) & = & -\frac{1}{N(t)}\frac{2}{\hbar}{\rm Im}\left[\sum_{\alpha}\mathcal{E}_{\alpha}{\rm Tr}\left[|\Psi_{\alpha}\rangle\langle\Phi_{\alpha}|\hat{\rho}(t)\right]\right]\nonumber\\
 & = & \sum_{\alpha}u_{\alpha}(t)\frac{\Gamma_{\alpha}}{\hbar}+2\sum_{\alpha}v_{\alpha}(t)\frac{E_{\alpha}}{\hbar}+\gamma_{0},\label{eq:main1}
\end{eqnarray}
where we introduce the coefficients $u_{\alpha},v_{\alpha}$ by
\begin{eqnarray}
u_{\alpha}(t) & \equiv & -\frac{1}{N(t)}{\rm Re}\left[{\rm Tr}\left[|\Psi_{\alpha}\rangle\langle\Phi_{\alpha}|\hat{\rho}(t)\right]\right]=-\frac{1}{N(t)}{\rm Tr}\left[\frac{|\Psi_{\alpha}\rangle\langle\Phi_{\alpha}|+|\Phi_{\alpha}\rangle\langle\Psi_{\alpha}|}{2}\hat{\rho}(t)\right],\\
v_{\alpha}(t) & \equiv & -\frac{1}{N(t)}{\rm Im}\left[{\rm Tr}\left[|\Psi_{\alpha}\rangle\langle\Phi_{\alpha}|\hat{\rho}(t)\right]\right]=-\frac{1}{N(t)}{\rm Tr}\left[\frac{|\Psi_{\alpha}\rangle\langle\Phi_{\alpha}|-|\Phi_{\alpha}\rangle\langle\Psi_{\alpha}|}{2}\hat{\rho}(t)\right].\label{eq:v}
\end{eqnarray}

Here, the density matrix $\hat{\rho}(t)$ appearing in $L(t)$ is originally defined as the Lindblad master equation in Eq. (\ref{eq:lind}).
On the other hand, since we consider noninteracting particles with one-body loss, the loss dynamics is fully characterized by a single-particle effective non-Hermitian dynamics.

From Eq. (\ref{eq:main1}), we infer that the long-time average of
the loss rate (subtracted by the offset $\gamma_{0}$) can be used to probe signatures of the PT-symmetry breaking:
\begin{equation}
\mathcal{O}(\gamma)\equiv\overline{\gamma_{0}-L(t)}\equiv\frac{1}{T}\int_{0}^{T}dt\left[\gamma_{0}-L(t)\right],
\end{equation}
where we choose $T$ to be $T\gg1/\gamma_{0}$. 
Above the threshold ($\gamma>\gamma_{\rm PT}$), there exists a conjugate pair of complex eigenvalues in the spectrum. Since a positive imaginary eigenvalue ($\Gamma_{\alpha}>0$) indicates an exponential lasing of the associated eigenmode, which turns out to be evident after the reduction of $L(t)$ to the language of single-particle dynamics, in the time regime $t>\gamma_{0}$ the dominant contribution in the sum of Eq. (\ref{eq:main1}) will come from the eigenstate $\alpha_{{\rm max}}$ whose imaginary part of the eigenvalue is maximum, resulting in the approximation
\begin{equation}\label{eq:ex1}
\mathcal{O}(\gamma)\propto\frac{\Gamma_{\alpha_{{\rm max}}}}{\hbar}>0\;\;\;\;\;\;\;(\gamma>\gamma_{\rm PT}).
\end{equation}
In particular, in the vicinity of the threshold, an imaginary part grows with the square-root scaling, and thus, we obtain $\mathcal{O}(\gamma)\propto\sqrt{\gamma-\gamma_{\mathcal{PT}}}$.

Meanwhile, below the threshold ($\gamma<\gamma_{{\rm PT}}$),
all the eigenvalues are real, and thus, the first term in Eq. (\ref{eq:main1}) is absent and there are no lasing eigenmodes found in the PT-broken regime. One can also check that the time average of $v_\alpha(t)$ should vanish, i.e., $\overline{v_{\alpha}(t)}\simeq0$ in the PT-unbroken regime. To see this, we  start from considering the simplest conditional density matrix evolving in the subspace of the initial total-atom number $N_0$
\begin{equation}
\hat{\tilde{\rho}}_{N_0}(t)=\hat{P}_{N_0}\hat{\rho}(t)\hat{P}_{N_0}=e^{-i\hat{H}_{{\rm eff}}t}\hat{\rho}(0)e^{i\hat{H}_{{\rm eff}}^{\dagger}t},
\end{equation}
where $\hat{P}_{N_0}$ denotes the projection operator on the subspace
of the atom number $N_0$. Then, we can calculate $v_{\alpha}(t)$ in this particular sector as
\begin{eqnarray}
 &  & -\frac{1}{N(t)}{\rm Im}\left[{\rm Tr}\left[|\Psi_{\alpha}\rangle\langle\Phi_{\alpha}|\hat{\tilde{\rho}}_{N_0}(t)\right]\right]\\
 & = & -\frac{e^{-N_0\gamma_{0}t}}{N(t)}{\rm Im}\left[{\rm Tr}\left[\sum_{\beta}e^{iE_{\beta}t}|\Phi_{\beta}\rangle\langle\Psi_{\beta}|\Psi_{\alpha}\rangle\langle\Phi_{\alpha}|e^{-iE_{\alpha}t}\hat{\tilde{\rho}}_{N_0}(0)\right]\right]\\
 & = & -\frac{e^{-N_0\gamma_{0}t}}{N(t)}\sum_{\beta\neq\alpha}{\rm Im}\left[e^{i(E_{\beta}-E_{\alpha})t}\langle\Psi_{\beta}|\Psi_{\alpha}\rangle{\rm Tr}\left[|\Phi_{\beta}\rangle\langle\Phi_{\alpha}|\hat{\tilde{\rho}}_{N_0}(0)\right]\right].\label{eq:pre2}
\end{eqnarray}
Crucially, in non-Hermitian systems  the left and right eigenvectors
constitute an orthonormal set while they themselves are in general nonorthogonal each other, i.e., $\langle\Psi_{\beta}|\Psi_{\alpha}\rangle\neq0$
for $\beta\neq\alpha$, leading to an oscillatory behavior
of the loss rate. When $E_{\beta}\neq E_{\alpha}$
if $\beta\neq\alpha$, the quantity~\eqref{eq:pre2} vanishes after taking the time average over a sufficiently long time.  Generalizing the argument to
other sectors of the atom numbers $N-1,N-2\ldots$, we can conclude that $\overline{v_{\alpha}(t)}\simeq0$, resulting in the relation in the PT-unbroken regime
\begin{equation}\label{eq:ex2}
\mathcal{O}(\gamma)\simeq0\;\;\;\;\;\;(\gamma<\gamma_{{\rm PT}}).
\end{equation}

\begin{figure}[t]
	\centering\includegraphics[width=150mm,bb=0 0 433 162]{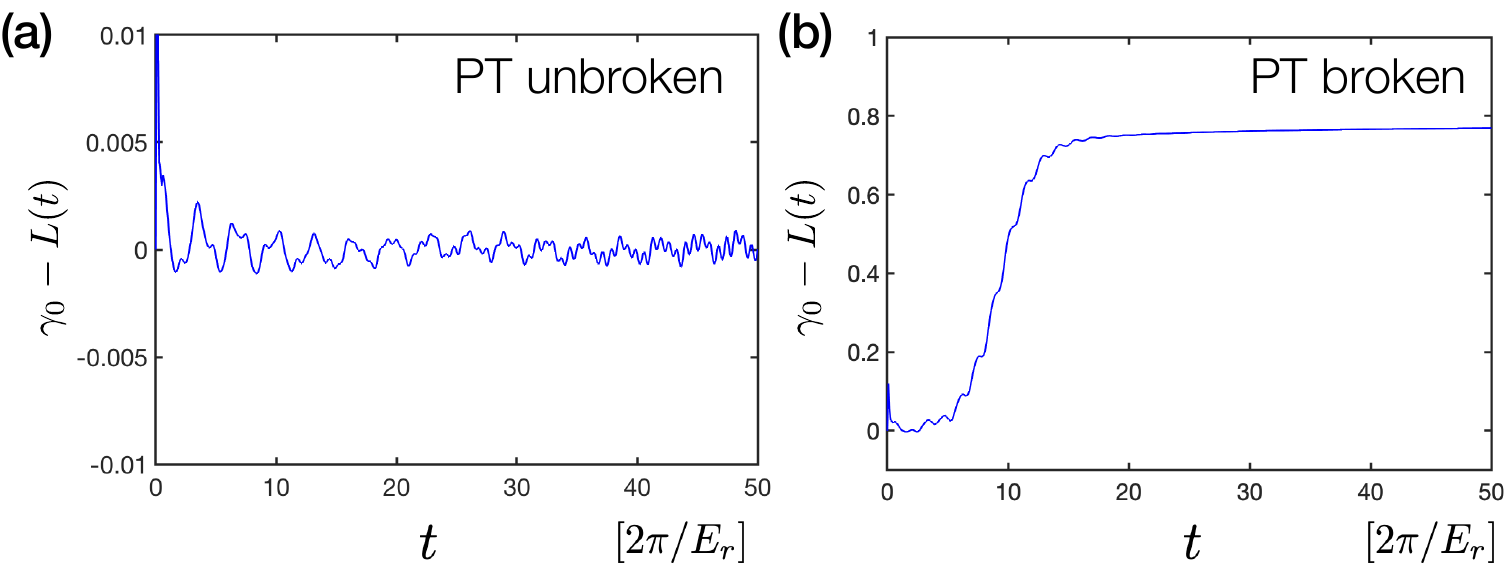}
	\caption{Numerical results on dynamics of the (negative) loss rate $\gamma_0-L(t)$ in (a) PT-unbroken regime at $\gamma=0.2$ and (b) PT-broken regime at $\gamma=0.5$. The initial state is assumed to be the Gibbs distribution of fermions $\hat{\rho}(0)\propto e^{-\beta\hat{H}}$ with $\beta=1/(k_{\rm 
	B}T)$, $T=0.5T_F$, and $\hat{H}$ being the Hamiltonian including an off-resonant periodic potential (but without dissipative terms). The dissipative optical potential is quenched at time $t=0$, and the density matrix is evolved in time according to the master equation~\eqref{eq:lind}. The time scale is plotted in terms of the recoil energy $E_r=\hbar^2\pi^2/(2md^2)$. We set the lattice depth to be $V_0=4E_r$.}
	\label{fig:num}
\end{figure}

Figure~\ref{fig:num} shows results of numerical simulations for time evolutions of the (negative) loss rate $\gamma_0-L(t)$  in (a) PT-unbroken and (b) PT-broken regimes. The results demonstrate the qualitative features in Eq.~\eqref{eq:ex1} and \eqref{eq:ex2} as expected from the above arguments; in the PT-unbroken regime, the  loss rate $\gamma_0-L(t)$ oscillates around zero in time due to nonorthogonality inherent to the non-Hermitian dynamics (Fig.~\ref{fig:num}(a)), while in the PT-broken regime it relaxes to a nonzero value corresponding to the maximum imaginary part of eigenvalues (Fig.~\ref{fig:num}(b)). We note that, since the Liouvillian of the master equation in the case of one-body loss is the quadratic operator, the density matrix during the time evolution can be exactly expressed as the Gaussian state when the initial state is an equilibrium state. 

\section{Experimental result}
We successfully realize a PT-symmetric quantum many-body system using a Bose-Einstein condensate (BEC) of  ${}^{174}\text{Yb}$ atoms in an optical superlattice.
One-body loss is introduced by exploiting rich internal degrees of freedom of Yb atoms.
This chapter describes the experimental schemes and results for one-body loss.

\subsection {Experimental setup}
\begin{figure}[b]
	\centering\includegraphics[width=6in]{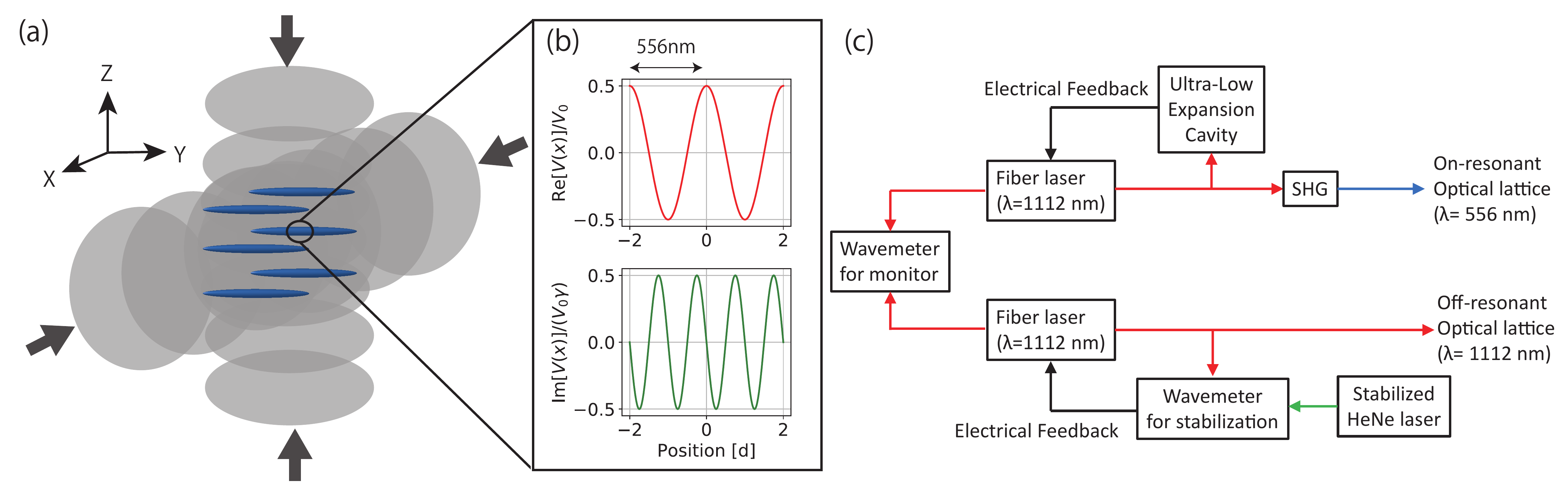}
	\caption{(a) Experimental setup of our lattice system.  Two-dimensional optical lattices (532 nm) are formed along the X and Z axes to create tight confinement potentials for 1D tubes. (b) Off-resonant (top) and on-resonant (bottom) lattice potentials with PT-symmetry. The superlattice of on-resonant and off-resonant lattices is formed along the Y axis. (c) Schematic diagram of frequency stabilization for the superlattice of on-resonant and off-resonant lattices. For the light source of the 556-nm and 1112-nm lattices, we used two fiber lasers operating at 1112 nm. One of the fiber lasers was used for the on-resonant lattice. The laser frequency was stabilized with an ultra low expansion cavity. Then 556 nm light was obtained with wave-guided second harmonic generation (SHG). Another fiber laser was used for the off-resonant lattice. The laser frequency was stabilized with a wavemeter. The frequency difference between the two fiber lasers was monitored with another wavemeter.}
	\label{fig:setup}
\end{figure}

Our experimental setup for the preparation of ${}^{174}\text{Yb}$ BEC and the basic optical lattice system are similar to our previous study~\cite{Tomita2017}.
In the present experiment, we create a two-dimensional (2D) array of one-dimensional (1D) tubes by tight confinement along the X and Z axes provided by a 2D optical lattice with 532 nm light (see Fig.~\ref{fig:setup} (a)).
We additionally introduce the superlattice which consists of two optical lattices generated by the retro-reflection of laser beams at 556 nm and 1112 nm along the Y axis (see Fig.~\ref{fig:setup} (b)).
Here we call the lattice generated by 556 nm light ``on-resonant lattice'' since the wavelength is resonant to the ${}^{1}\text{S}_0$-${}^{3}\text{P}_1$ transition of Yb atoms and the lattice by 1112 nm ``off-resonant lattice'' since it does not correspond to any resonance.
To realize PT-symmetry of the system, the relative phase between the on-resonant and off-resonant lattices is finely tuned.
The on-resonant lattice provides the excitation of the atoms in the ground state ${}^{1}\text{S}_0$ of our interest to the excited ${}^{3}\text{P}_1$ state, naturally introducing one-body loss in the system.
However, to prevent an unnecessary heating effect, it is also required that the excited atom should not return to the ground state, as is assumed in the theory.
For this purpose, we exploited the repumping technique, which is described in detail in the next subsection.

\subsection{Repumping}
\begin{figure}[t]
	\centering\includegraphics[width=6in]{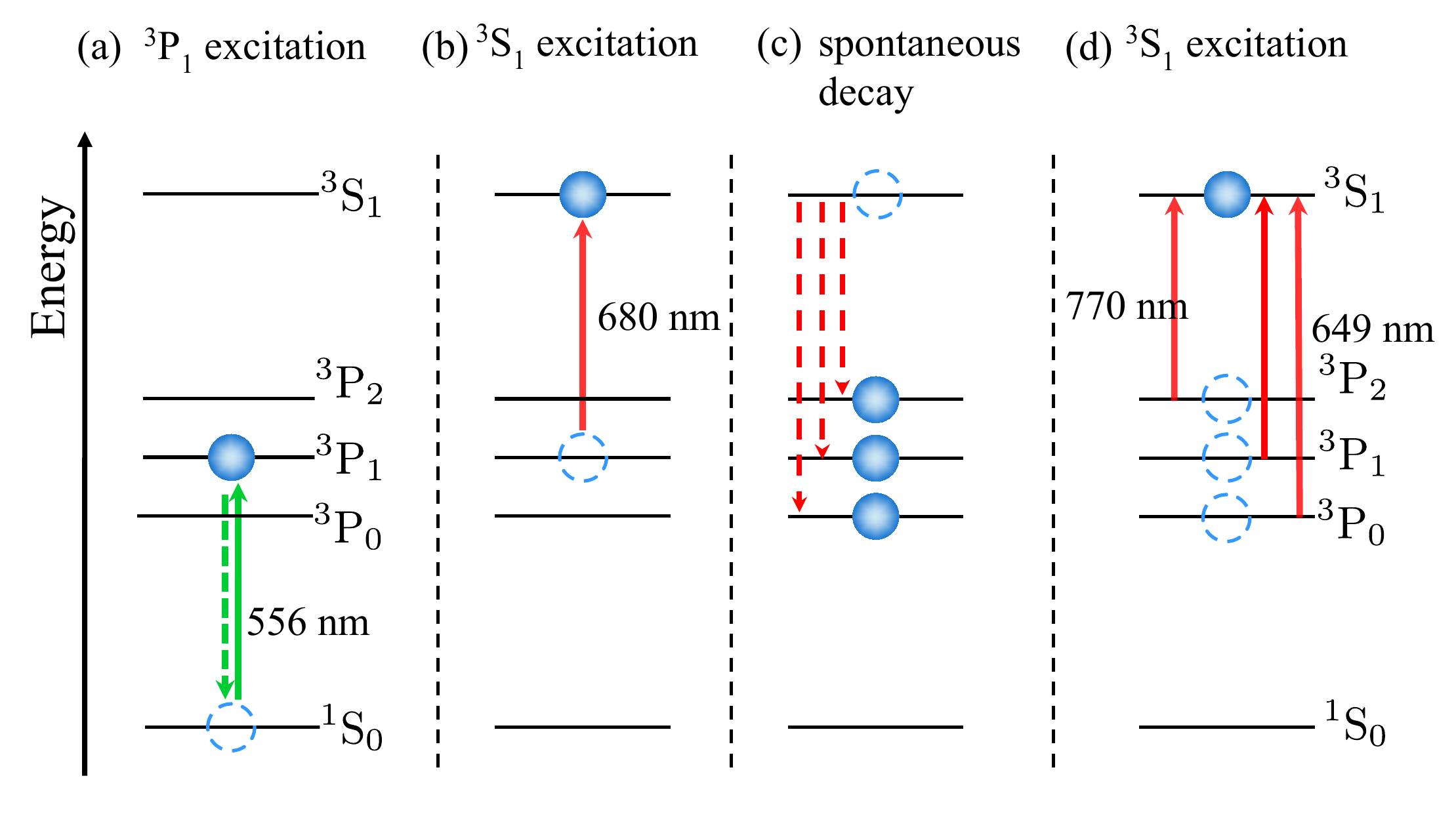}
	\caption{Repumping scheme for one-body loss. (a) Atoms in the ground state ${}^{1}\text{S}_0$ are excited to the ${}^{3}\text{P}_1$ state. (b) Atoms in the ${}^{3}\text{P}_1$ state are excited to the ${}^{3}\text{S}_1$ state, rather than decay to the ground state. (c) Atoms in the ${}^{3}\text{S}_1$ state spontaneously decay to one of the ${}^{3}\text{P}_2$, ${}^{3}\text{P}_1$, and ${}^{3}\text{P}_0$ states. (d) All atoms are excited again to the ${}^{3}\text{S}_1$ state.}
	\label{fig:repump}
\end{figure}

Yb atoms excited in the ${}^{3}\text{P}_1$ state decay into the ${}^{1}\text{S}_0$ ground state by spontaneous emission.
As a result, the atom obtains a recoil energy.
The single recoil energy is, however, not enough for the atoms to escape from a deep optical lattice like in our experiment (see 3.3 for detail).
It is also true that, because the photon absorption and emission constitutes a closed cycle, the atom escapes from the trap after the total energy of the atom becomes high until it reaches the depth of the trap potential.
This causes serious heating in the system, which is quite unfavorable for our purpose.
In order to realize the successful removal of an atom after a single excitation, we additionally use the ${}^{3}\text{P}_1$-${}^{3}\text{S}_1$ excitation (see Fig.~\ref{fig:repump}).
The ${}^{3}\text{P}_1$-${}^{3}\text{S}_1$ transition is the strong dipole-allowed transition, and if the Rabi frequency of the ${}^{3}\text{P}_1$-${}^{3}\text{S}_1$ transition is taken be quite large compared to the one of the ${}^{1}\text{S}_0$-${}^{3}\text{P}_1$ transition, the atoms in the ${}^{3}\text{P}_1$ are more likely to be excited into the ${}^{3}\text{S}_1$ state than to undergo a spontaneous decay to the ${}^{1}\text{S}_0$ ground state.
Consequently, the excited atoms in the ${}^{3}\text{S}_1$ state will spontaneously decay into any one of the ${}^{3}\text{P}_2$, ${}^{3}\text{P}_1$, and ${}^{3}\text{P}_0$ states.
The atom in the ${}^{3}\text{P}_1$ state will be excited into the ${}^{3}\text{S}_1$ state again.
In addition, the atom in the ${}^{3}\text{P}_2$ (${}^{3}\text{P}_0$) state will be excited into the ${}^{3}\text{S}_1$ state by additional ${}^{3}\text{P}_2$-${}^{3}\text{S}_1$ (${}^{3}\text{P}_0$-${}^{3}\text{S}_1$) transition.
As a result, once an atom in the ground state is excited into the ${}^{3}\text{P}_1$ state, the atom is removed from the trap without going back to the ground state.
The typical Rabi frequency in our experiment is 1 MHz (60 MHz) for the ${}^{1}\text{S}_0$-${}^{3}\text{P}_1$ (${}^{3}\text{P}_1$-${}^{3}\text{S}_1$) excitation.
Therefore the effective two-level model with one-body loss is realized.
In the following we refer to the three light beams resonant to the ${}^{1}\text{S}_0$-${}^{3}\text{P}_x$ ($x=0,1,2$) transitions as repumping beams.

\begin{figure}[t]
	\centering\includegraphics[width=6in]{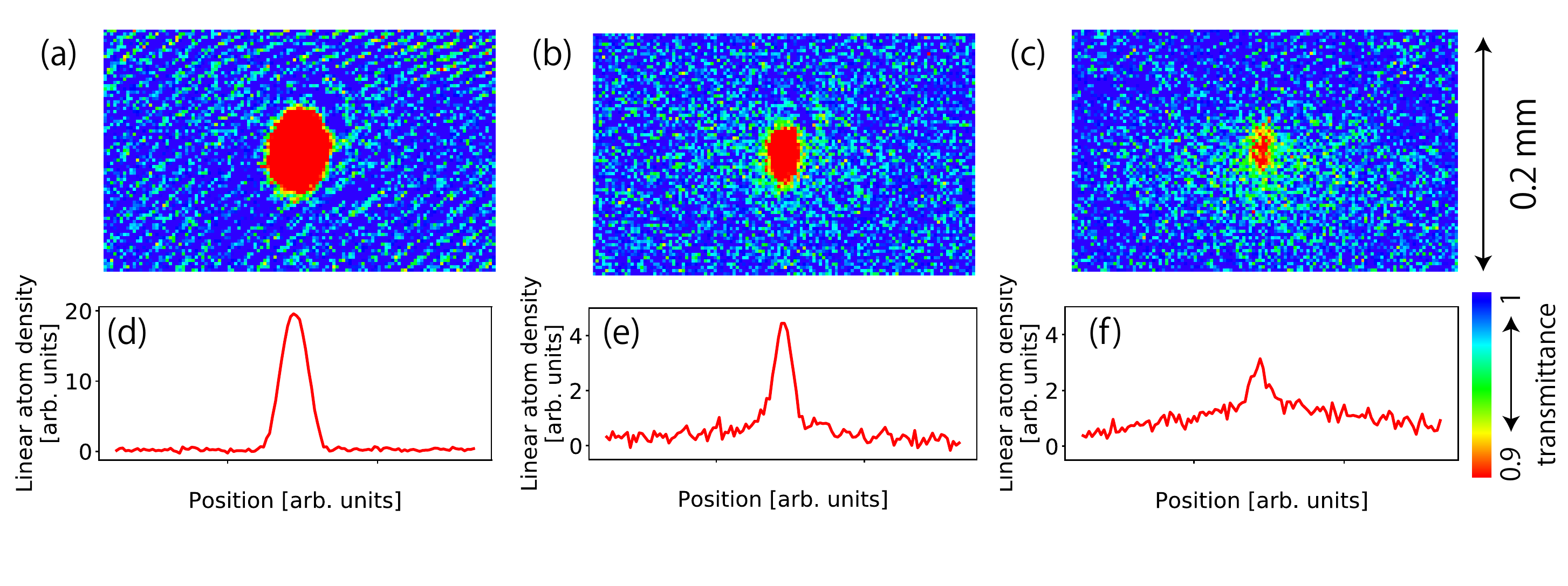}
	\caption{(a)-(c) Absorption images of atoms after 16 ms Time-of-Flight. (a) Resonant ${}^{1}\text{S}_0$-${}^{3}\text{P}_1$ light is not irradiated. (b) The resonant light is applied for 1.5 ms with simultaneous application of repumping beams. (c) The resonant light is applied for 1.5 ms without repumping beams. (d)-(f) Atom linear density integrated along the vertical direction. (d)-(f) correspond to (a)-(c), respectively.}
	\label{fig:one-body-repump}
\end{figure}
Absorption images after 16 ms Time-of-Flight (ToF) reveal the effects of the repumping beams.
Figure~\ref{fig:one-body-repump}(a) shows a typical BEC ToF signal obtained with no ${}^{1}\text{S}_0$-${}^{3}\text{P}_1$ resonant light irradiation.
In the presence of ${}^{1}\text{S}_0$-${}^{3}\text{P}_1$ resonant light irradiation, the atom numbers are reduced, as is shown in Figs.~\ref{fig:one-body-repump}(b) and (c).
When no repumping beams are applied, considerable thermal components are accompanied as is shown in Fig.~\ref{fig:one-body-repump}(c).
In contrast, owing to the simultaneous application of repumping beams, the heating effect is significantly reduced, as shown in Fig.~\ref{fig:one-body-repump}(b), which indicates that the atoms are removed from the trap without going back to the ground state in the presence of the repumping beams.

\subsection{Relative phase measurement}
The relative phase $\phi$ between the on-resonant and off-resonant lattices should be finely tuned in order to satisfy the PT-symmetry condition, which is essentially important in the present experiment.
In order to determine the relative phase, we measure the dependence of the atom loss on the relative phase.
\begin{figure}[t]
	\centering\includegraphics[width=6in]{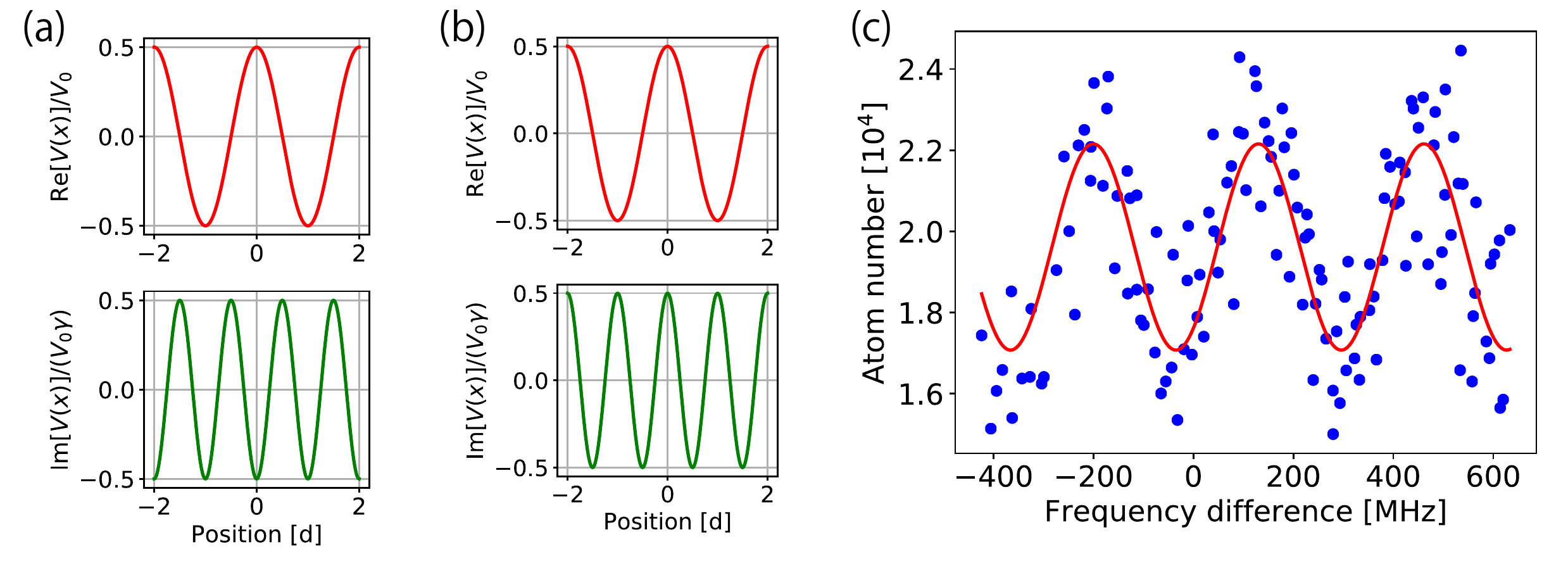}
	\caption{(a),(b) Off-resonant (top) and on-resonant (bottom) lattice potentials. Relative phases $\phi$ are (a) $+\pi/2$ and (b) $-\pi/2$, respectively (c) Remained atom number as a function of frequency difference of the on-resonant lattice and the off-resonant lattice. The solid line shows the fitting result using a sinusoidal function.}
	\label{fig:one-body-scan}
\end{figure}
In our experiment the off-resonant lattice is deep enough so that trapped atoms are localized in each site.
In this case, the behavior of atom loss by the on-resonant lattice depends on the relative phase.
When the relative phase is $\phi =+\pi/2$ ($\phi =-\pi/2$), atom loss has a maximum (minimum) value (see Fig.~\ref{fig:one-body-scan} (a) and (b)). 

In order to measure the dependence of atom loss on the relative phase, we performed the following experiment.
After preparation of BEC in an optical trap, the optical lattices were adiabatically turned on until the depth of ($V_x$, $V_y$, $V_z$)=($18E_{R, 532\text{nm}}$, $8E_{R,1112\text{nm}}$, $15 E_{R, 532\text{nm}}$), where $V_i$ ($i=x,y,z$) are the lattice depths along the $i$ direction, and $E_{R, \lambda}=h^2/(2m\lambda^2)$ is a recoil energy.
Here, atoms form a  Mott insulator state in the center of the trap.
Then the on-resonant lattice and repumping beam are simultaneously applied during 0.5 ms and the remained atom number was measured by absorption imaging.
The relative phase is controlled by changing the frequency difference between the fundamental light for on-resonant lattice (556 nm) before the second-harmonic generation and off-resonant lattice (1112 nm).
Figure~\ref{fig:one-body-scan}(c) shows the remained atom number as a function of frequency difference.
Clear periodic dependence is observed, from which we can determine the phase at which PT-symmetric condition ($\phi=0$) is satisfied.
This clear dependence did not appear without the repumping beams.
In this way, we successfully realize a PT-symmetric quantum many-body system in a well controllable manner.

\subsection{Loss rate measurement}
\begin{figure}[t]
	\centering\includegraphics[width=6in]{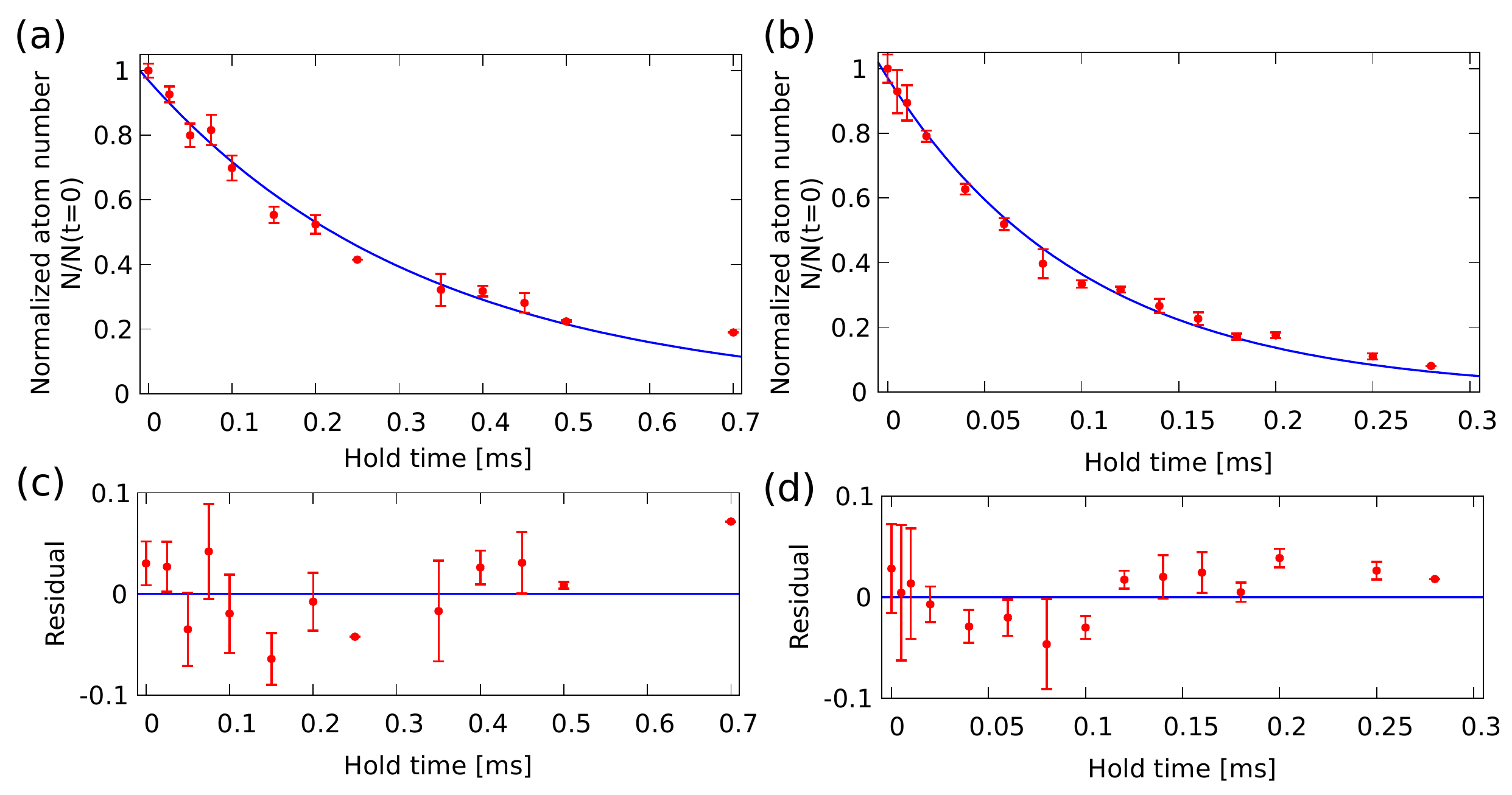}
	\caption{(a),(b) Remained atom number as a function of hold time with (a) small or (b) large one-body loss. The solid lines show exponential fitting. From the fits, we estimated the loss rate $\gamma$ to be (a) $2.9$ or (b) $9.6$ kHz. (c),(d) Residuals of the fits in the cases of (a),(b), respectively.}
	\label{fig:one-body}
\end{figure}

As we discuss in the theoretical framework, the atom loss rate is an important quantity to characterize the phase of the PT-symmetric system.
To measure loss behavior, we first adiabatically turned on the two 532-nm optical lattices along the X and Z directions as well as an off-resonant lattice along the Y direction up to ($V_x$, $V_y$, $V_z$)=($18E_{R, 532\text{nm}}$, $5 E_{R, 1112\text{nm}}$, $15 E_{R, 532\text{nm}}$).
Then we suddenly turned on the on-resonant lattice, and repump beams.
The relative phase was stabilized to the PT-symmetric condition point.
Figure~\ref{fig:one-body} shows remained atom numbers as a function of hold time on the PT-symmetric lattice.
The solid lines are fitting results with the exponential function
\begin{equation}
N(t)=N_0 \exp(-\gamma' t).
\end{equation}
Figure~\ref{fig:one-body}(a) shows a weak-loss case.
From the exponential fitting of remained atom number, we estimated the loss rate $\gamma'$ to be $2.9$($1$) kHz.
At this parameter, PT symmetry is not broken and the observed decay is well fitted by an exponential decay due to the constant decay term represented by the last term of Eq.(\ref{eq:non-her}).

Figure~\ref{fig:one-body}(b) shows a strong-loss case.
The light intensity of the ${}^{1}\text{S}_0$-${}^{3}\text{P}_1$ transition is about 3.3 times higher than that of the Fig.~\ref{fig:one-body}(a) case.
At this parameter, PT symmetry is expected to be broken between the first excited band and the second excited band in the non-interacting case.
The atom loss was well fitted by an exponential decay.
While the additional loss is the signature of PT-broken phase, the loss rate $\gamma'$ was $9.6$($3$) kHz which corresponds to 3.3 times of the loss rate observed at the weak-loss case in Fig.~\ref{fig:one-body}(a) within the error bars.
This means that the current measurement uncertainty is too large to detect the expected transition between PT-unbroken and PT-broken phases.

\section{Discussion and Conclusion}
We developed the theoretical framework and experimental setup for the PT-symmetric lattice using one-body loss.
As a result, we successfully create the PT-symmetric non-Hermitian many-body system using ultracold Bose gas.
Atom loss behaviors we measured were well described by using usual one-body atom loss models.

In order to observe the unique behavior of the PT-broken and unbroken transition with respect to atom loss, the stability and sensitivity of atom number measurement should be improved.
We used here absorption imaging method and it is difficult to measure atom numbers smaller than $10^3$ atoms because of noise on images originated from interference fringes.
We already developed another imaging method, fluorescence imaging: atoms are recaptured by using a magneto-optical trap (MOT) with the ${}^1\text{S}_0$ - ${}^1\text{P}_1$ transition, and the fluorescence from the MOT is measured~\cite{Kato2012, Kato2016, Takasu2017}.
We already confirmed that we could measure atom number of about $10^2$ Yb atoms with accuracy of about $10$ atoms~\cite{Kato2012, Takasu2017}.

Another possible method is to create an ultracold atom system with a {\it gain} term.
Simply thinking, it is hard to include the gain term by injection of atoms from environment.
However, effective gain factor with anti-magic lattice in a two-orbital system is theoretically suggested~\cite{Gong2018}.
In addition, an effective atom gain term by use of tilted lattice is theoretically suggested~\cite{Kreibich2014}.

When operated in the presence of interaction, the experimental system developed here should allow one to study novel many-body phenomena unique to non-Hermitian regimes such as anomalous quantum phase transitions and criticality \cite{YA17nc,LAS18,NM18,MN19}. 
Even in noninteracting regimes, unique nonequilibrium features such as the violation of the Lieb-Robinson bound \cite{YA18,DB19} and the unconventional Kibble-Zurek mechanism \cite{YS17,DB192} should be investigated with the help of single-atom-resolved measurement techniques.  

\section*{Acknowledgment}
This work was supported by the Grant-in-Aid for Scientific Research of the Ministry of Education, Culture Sports, Science, and Technology / Japan Society for the Promotion of Science (MEXT/JSPS KAKENHI) Nos. JP25220711, JP17H06138, JP18H05405, JP18H05228, and JP19K23424; the Impulsing Paradigm Change through Disruptive Technologies (ImPACT) program; Japan Science and Technology Agency CREST (No. JPMJCR1673), and MEXT Quantum Leap Flagship Program (MEXT Q-LEAP) Grant Number JPMXS0118069021.
Y.A. acknowledges support from the Japan Society for the Promotion of Science through KAKENHI Grant No.~JP16J03613.
R.H. was supported by the Japan Society for the Promotion of Science through Program for Leading Graduate Schools (ALPS) and JSPS fellowship (JSPS KAKENHI Grant No. JP17J03189). 
Y.K. was supported by JSPS fellowship (JSPS KAKENHI Grant No.JP17J00486).

\appendix

\setcounter{equation}{0}
\renewcommand{\theequation}{A\arabic{equation}}
\setcounter{figure}{0}
\renewcommand{\thefigure}{A\arabic{figure}}
\renewcommand{\thesubsection}{A.\arabic{subsection}}

\section{PT-symmetric non-Hermitian quantum many-body system with two-body loss}
\subsection{Theoretical description}
Photoassociation can be utilized for introducing a two-body loss~\cite{Tomita2017}.
In the present setup, a photoassociation resonance is available near atomic ${}^{1}\text{S}_0$ - ${}^{3}\text{P}_{1}$ resonance.

We here briefly discuss the case in which light induces two-body loss of atoms, rather than one-body loss discussed in the main text. Following the same procedure as done for the one-body case, one can arrive at the master equation of the ground-state atoms \cite{JJGR09,DSG09} 
\begin{eqnarray}
\frac{d\hat{\rho}}{dt}&\!=\!&-\frac{i}{\hbar}(\hat{H}_{{\rm eff}}\hat{\rho}-\hat{\rho}\hat{H}_{{\rm eff}}^{\dagger})+\int dx\frac{|\Omega({\bf x})|^{2}}{\Gamma}\hat{\Psi}({\bf x})\hat{\Psi}({\bf x})\hat{\rho}\hat{\Psi}^{\dagger}({\bf x})\hat{\Psi}^{\dagger}({\bf x}),\\
\hat{H}_{{\rm eff}}&\!=\!&\int d{\bf x}\hat{\Psi}^{\dagger}({\bf x})\left(-\frac{\hbar^{2}\nabla^{2}}{2m}\!+\!U({\bf x})\right)\hat{\Psi}({\bf x})\!-\!\int d{\bf x}\frac{i|\Omega({\bf x})|^{2}}{2\Gamma}\hat{\Psi}^{\dagger}({\bf x})\hat{\Psi}^{\dagger}({\bf x})\hat{\Psi}({\bf x})\hat{\Psi}({\bf x}).
\end{eqnarray}
We note that the non-Hermitian term corresponding to two-body loss of atoms can be expressed as a purely imaginary interaction term, and thus the system  becomes an intrinsically interacting many-body problem even when $u=0$, where $u$ is defined in Eq.(\ref{eq:u}). As a result, the density matrix during the dynamics in general deviates from the Gaussian limit and cannot be reduced to a single-particle non-Hermitian dynamics, making contrast to the case of one-body loss.

Here we are interested in the case that the off-resonance lattice is so deep that a tight-binding approximation is valid and that atoms are in the first band.
In addition, we consider Hubbard-like approximation, that is, the atom in each site is well described by a Wannier function of non-interacting and non-dissipation case.
The Wannier function in the first band is symmetric and the dissipation lattice is anti-symmetric with respect to a central position of a lattice site.
When the phase shift of the dissipation lattice is set to be $\phi=0$ (i.e., the PT-symmetric case), we obtain
\begin{equation}
\sum_i\left(\int dx \sin\left(\frac{4\pi x}{d}\right)|w_i(x)|^4\right)\hat{a}_i^{\dagger}\hat{a}_i^{\dagger}\hat{a}_i\hat{a}_i=0.
\end{equation}
Thus, the spatial dependence in the anti-Hermitian part of $H_{\rm eff}$ does not play any significant roles; the dynamics of $H_{\rm eff}$ reduces to the usual Hermitian one aside the normalization of the wavefunction due to the homogeneous loss. In contrast, if the off-resonant lattice is weak, the spatial dependence comes in, and the dynamics is governed by the PT-symmetric non-Hermitian operator $H_{\rm eff}$, which is genuinely distinct from a Hermitian one.
We focus here on Tonks\textendash Girardeau gas, which is a many-body system of strongly interacting Bose gas confined in tight 1D tubes, and behaves as a free fermi gas~\cite{Paredes2004}.

\subsection {Experimental setup}
\begin{figure}[tb]
	\centering\includegraphics[width=3in]{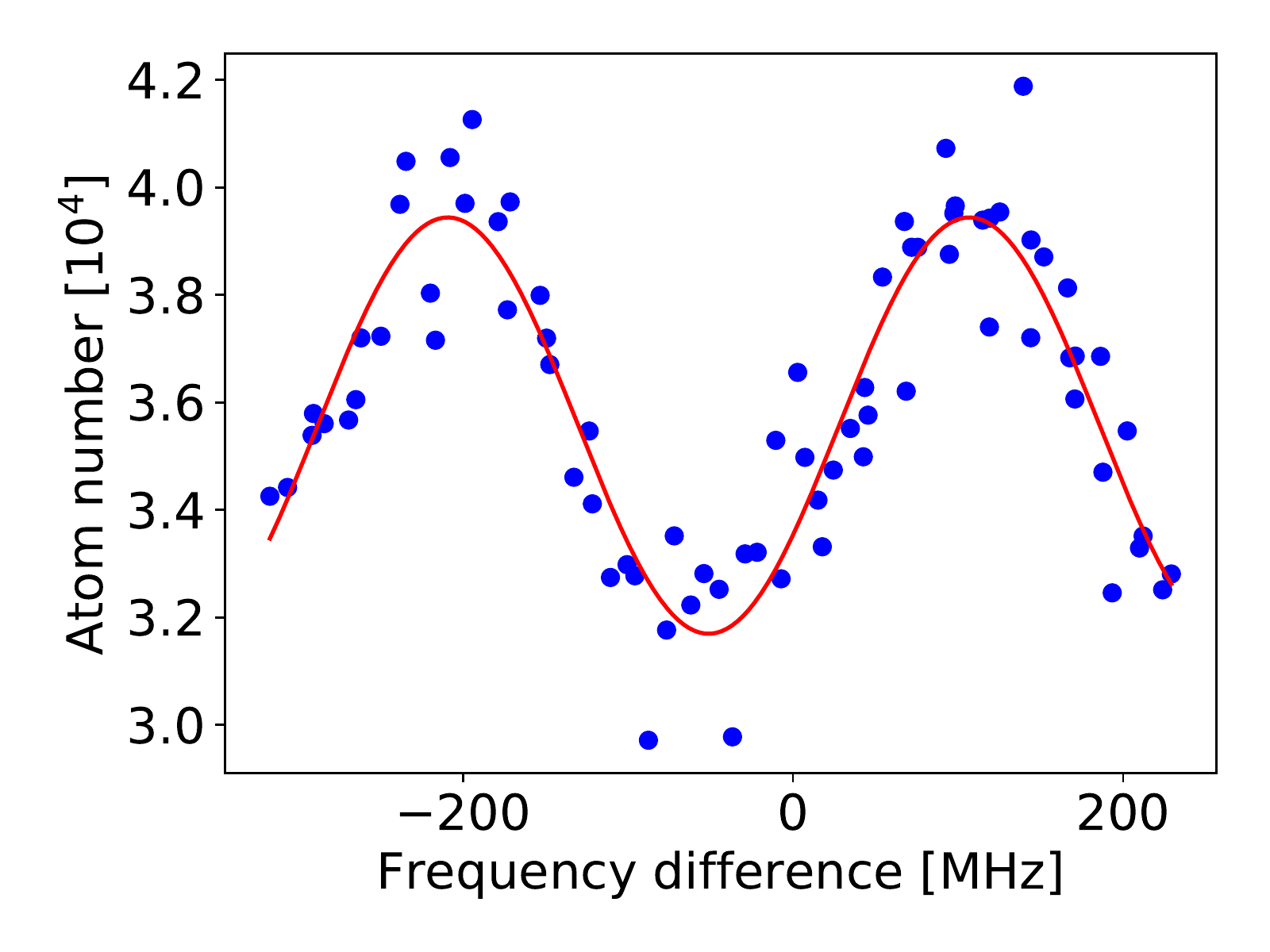}
	\caption{Remained atom number as a function of frequency difference of the on-resonant lattice and the off-resonant lattice. The solid line shows the fitting result using a sinusoidal function.}
	\label{fig:two-body-scan}
\end{figure}

The experimental setup is almost the same as the one-body loss case (see the main text), except the following two points.
First is about the repumping lasers.
Photoassociation is an optical transition from two isolated atoms into one excited diatomic molecule.
Then the excited molecule decays by spontaneous emission into two isolated atoms or one molecule in the ground state.
In our experimental condition, the two atoms are quickly converted to one molecule. 
Therefore, the trap loss is realized by solely applying the photoassociation beam without need of repumping beams, different from one-body loss.
The relative-phase measurement clearly demonstrates this feature.
The procedure is the same as the one-photon loss case.
The result is shown in Fig.~\ref{fig:two-body-scan}.
It is noted that there are no repump beams in this measurement, which shows that atoms are removed without heating.

Second is the laser frequency of the dissipation lattice.
We used a photoassociation resonance which is red-detuned ($\sim 10$ MHz) from the atomic ${}^{1}\text{S}_0$-${}^{3}\text{P}_1$ resonance.
The dissipation lattice also created non-zero potential because of non-zero detuning from the atom resonance.
We measured the potential depth, which was order of recoil energy and was not negligible.
In order to  cancel out the potential, we superimposed an additional 556-nm lattice, which is far red-detuned (337.6 MHz) from the atomic ${}^{1}\text{S}_0$-${}^{3}\text{P}_1$ resonance (in the following we refer to this lattice as ``compensation lattice'').
The detuning is so large that additional atom loss is negligible.
It is noted that the frequency difference between the on-resonant lattice and compensation lattice is 327.6 MHz and is obtained from the fitting result shown in Fig.~\ref{fig:two-body-scan}.
In order to tune the intensity of the compensation lattice, we measure the sum of lattice potential of the on-resonant lattice and compensation lattice with a pulsed-lattice method.

\subsection{Atom preparation}
First we measured a Lieb-Liniger parameter without both of the off-resonant and the on-resonant lattices.
We prepared $2.0 \times 10^4$ ${}^{174}\text{Yb}$ BEC atoms in an optical trap.
We adiabatically turned on the two 532-nm optical lattices along the X and Z directions
to ($V_x$, $V_z$)=($18E_{R, 532\text{nm}}$, $15 E_{R, 532\text{nm}}$).
The atoms were trapped in 1D tubes and we estimated a Lieb-Liniger parameter $\gamma_{LL}$ by measuring photoassociation rate and $\gamma_{LL}$ is about 1.54.
Therefore, trapped atoms satisfied Tonks-Giradeau gas condition $\gamma_{LL}>1$.

\subsection{Loss measurement}
\begin{figure}[t]
	\centering\includegraphics[width=6in]{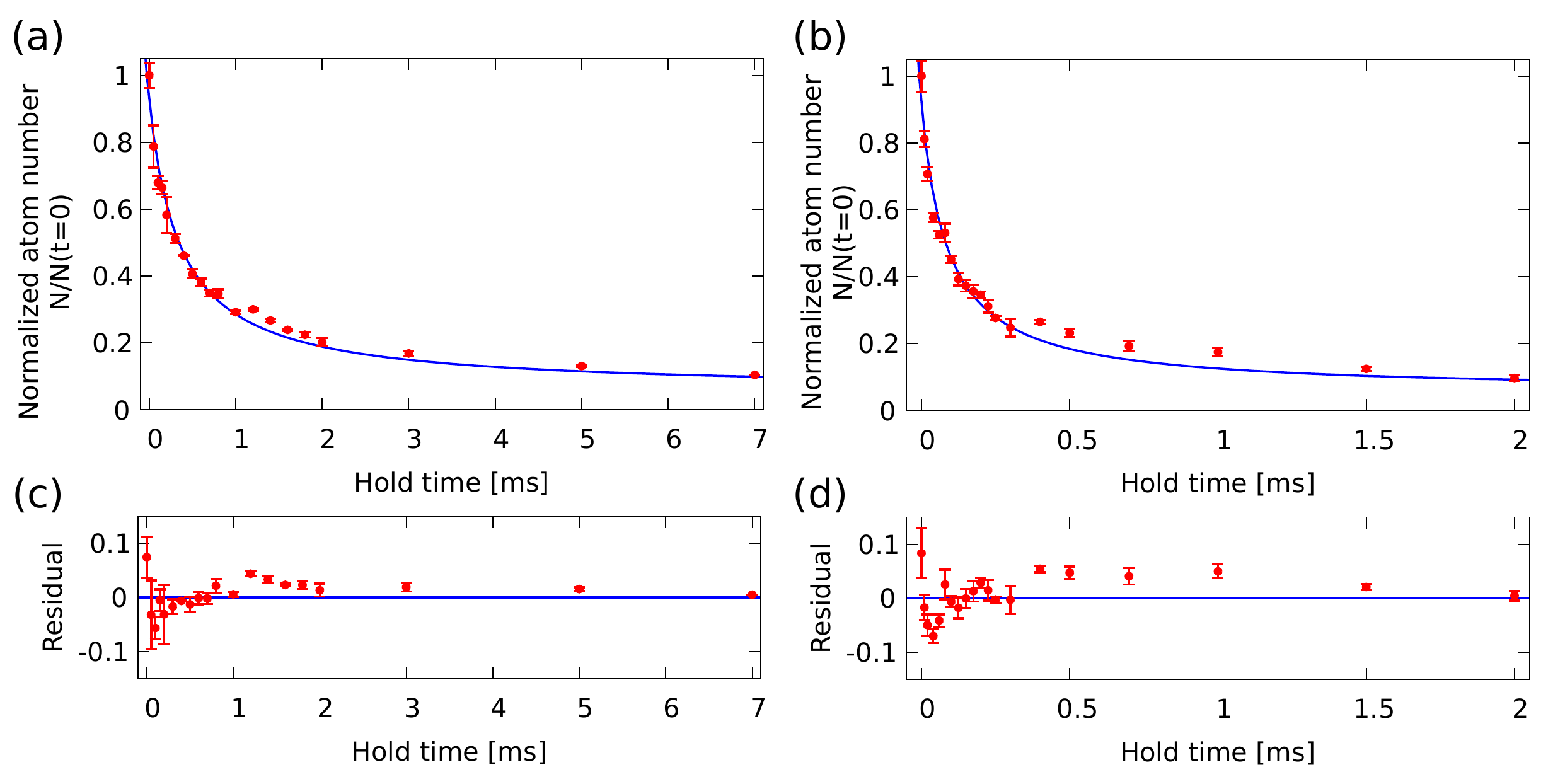}
	\caption{(a),(b) Remained atom number as a function of hold time with (a) small or (b) large two-body loss. The solid lines show the two-body-loss fitting using Eq.~(A4). From the fit, we estimated the loss rate $\beta$ to be (a) $2.0$ or (b) $13$ kHz.  (c),(d) Residuals of the fits in the cases of (a),(b), respectively.}
	\label{fig:two-body}
\end{figure}

Loss measurement was performed by almost the same procedure as the one-photon loss case.
After preparation of ${}^{174}\text{Yb}$ BEC in an optical trap, we adiabatically turned on the two 532-nm optical lattices along the X and Z directions to ($V_x$, $V_y$, $V_z$)=($18E_{R, 532\text{nm}}$, $3 E_{R, 1112\text{nm}}$, $18 E_{R, 532\text{nm}}$).
Then we suddenly turned on the on-resonant lattice and compensation lattice.
The relative phase was stabilized to the PT-symmetric condition point.
Figures~\ref{fig:two-body} (a) and (b) show remained atom numbers as a function of hold time in the case of the PT-symmetric lattice.
The solid lines show fitting results with two-body loss model
\begin{equation}
N(t)=\frac{N_0}{1+\beta t}+b. \label{eq:twobody}
\end{equation}
Figure~\ref{fig:two-body}(a) shows a weak-loss case.
From the fitting of remained atom number, we estimated the loss rate $\beta$ to be $2.0$($2$) kHz.
Figure~\ref{fig:two-body}(b) shows a strong-loss case.
The photoassociation light intensity is about 7.5 times higher than that of the Fig.~\ref{fig:two-body}(a) case.
The loss rate $\beta$ was $13$($1$) kHz which almost corresponds to 7.5 times of the loss rate observed at the weak-loss case in Fig.~\ref{fig:two-body}(a).
Again this means that the current measurement uncertainty is too large to discuss the PT-unbroken and PT-broken transition.


%


%

\end{document}